\documentclass{emulateapj}
\usepackage{natbib}
\shorttitle{}
\usepackage{color}
\usepackage{amsmath}
\usepackage{epstopdf}
\def\kepler{{\slshape Kepler}}
\begin{document}
\title{}
\author{John Moriarty}
\affil{Yale University, New Haven, CT 06511, USA}
\email{john.c.moriarty@yale.edu}
\and
\author{Sarah Ballard}
\affil{Massachusetts Institute of Technology, Cambridge, MA 02139 USA}

\title{The Kepler Dichotomy in Planetary Disks: Linking Kepler Observables to Simulations of Late-Stage Planet Formation}

\begin{abstract}

NASA’s \kepler\ Mission uncovered a wealth of planetary systems, many with planets on short-period orbits. These short-period systems reside around 50\% of Sun-like stars and are similarly prevalent around M dwarfs. Their formation and subsequent evolution is the subject of active debate. In this paper, we simulate late-stage, in-situ planet formation across a grid of planetesimal disks with varying surface density profiles and total mass. We compare simulation results with observable characteristics of the \kepler\ sample. We identify mixture models with different primordial planetesimal disk properties that self-consistently recover the multiplicity, period ratio and duration ratio distributions of the \kepler\ planets. We draw three main conclusions:  (1) We favor a ``frozen-in" narrative for systems of short period planets, in which they are stable over long timescales, as opposed to metastable. (2) The ``\kepler\ dichotomy", an observed phenomenon of the \kepler\ sample wherein the architectures of planetary systems appear to either vary significantly or have multiple modes, can naturally be explained by formation within planetesimal disks with varying surface density profiles. Finally, (3) we quantify the nature of the ``\kepler\ dichotomy" for both GK stars and M dwarfs, and find that it varies with stellar type. While the mode of planet formation that accounts for highly multiplistic systems occurs in 24$\pm$7\% of planetary systems orbiting GK stars, it occurs in 63$\pm$16\% of planetary systems orbiting M dwarfs.
\end{abstract}

\keywords{planets and satellites: formation}
%%%%%%%%%%%%%%%%%%%%%%%%%%%%%%%%%%%%%%%%%%%%%%%%%%%%%%%%
\section{Introduction}
 One of the most important findings of NASA's \kepler\ mission is the prevalence of planets on relatively short period orbits. M dwarfs host, on average, 2.5 planets per star with radii between 1 and 4$R_\oplus$ and orbital periods shorter than 200 days \citep{Dressing15b, Morton14b} and nearly half of all G and K stars host at least 1 planet within the same range \citep{Petigura13, Silburt15}. The ubiquity of these planetary systems has sparked a number of investigations of their underlying architecture. The most straight-forward observable to compare with predictions is the number of transiting planets (``tranets") per system. Constructing a set of synthetic planetary systems and then ``observing" them from random angles provides a means of comparison to the \kepler\ tranet yield. These studies constrain the underlying multiplicity of the systems and distribution of planet inclinations. The general consensus from these simulations is that planetary systems tend to be very flat (mutual inclinations less than a few degrees) \citep{Fang12b, Swift13, Ballard14, Fabrycky14, Figueira12}.

 However, there exists a strong degeneracy between the underlying multiplicity and the mutual inclinations of the planets. That is, systems with more planets with higher mutual inclinations will present the same number of transiting planets as systems with fewer planets with lower mutual inclinations. Transit duration ratios of adjacent planets, which are sensitive to the inclination distribution of planets, provide a means to break this degeneracy. Both \citet{Fang12b} and \cite{Fabrycky14} folded transit duration ratios into their analyses and concluded that most systems have 1-2 planets with mutual inclinations between planets less than 3$^\circ$.

 Other studies have found that the observed population of planets is not well described by a single-component model of system architecture. \citet{Lissauer11} and \citet{Ballard14} both find that their best fitting single component models significantly under-predict the number of singly transiting systems. \citet{Hansen13} compare planetary systems created in late stage planet formation simulations to observed systems and similarly find that singly transiting systems are 50\% more common in the observed population than their modeled population. Looking beyond multiplicity, \citet{Xie14} compared the fraction of transiting planets showing transit timing variations (TTVs) for single vs multi-planet systems and found that systems with four or more transiting planets show TTVs at a rate $\sim$5 times higher than those of singly transiting systems. These findings all suggest that the underlying architecture of planetary systems is better described by either a mixture model or a continuum of architectures.

 Qualitatively, a mixture model must consist of one component with many planets on low mutual inclinations that accounts for the majority of systems with two or more transiting planets. The other component must account for the majority of systems with only one transiting planet. These latter systems either host fewer planets, or their planets possess high mutual inclinations so that only one planet will transit from any given viewing angle. The utility of transit duration ratios as a tool for probing mutual inclination is lost for systems with only one transit. \citet{Morton14} find that the stellar obliquity of systems with only one transiting planet are systematically higher than for systems with multiple transiting planets. This may suggest that the second population is made up of planets with high mutual inclinations. However, it is possible that whichever potential mechanism reduces the underlying multiplicity of systems also increases obliquity.

 A number of authors have explored the possibility that a dichotomy in the structure of planetary systems can arise from the long term evolution of dynamically full systems. \citet{Volk15} suggest that systems of many tightly packed planets are the normal outcome of planet formation and that some fraction of these systems eventually undergo a dynamical instability which leads to the consolidation of these systems into fewer planets. Similarly, \citet{Pu15} find that many of the \kepler\ multi-planet systems are metastable. They propose that the lower multiplicity systems are resultant from more densely packed systems that have undergone a period of dynamical instability. On the other hand, \citet{Johansen12} find that dynamical instability is probably not the cause of the \kepler\ dichotomy. They followed the evolution of triple systems with the goal of producing doubles with high mutual inclinations. However, they found that the timescales needed for instability to occur are very long unless the planets are very high mass. Planets of 0.1 to 10 Jupiter masses are needed for instability to occur within 10 billion years. Similarly \citet{Becker15} found that self-excitation is rarely sufficient to remove planets from the transiting geometry.

 We explore this hypothesis through N-body simulations of late stage planet formation and a comparison of these simulation results to the observed population of planets. We investigate whether two properties: (1) the total mass of planetesimals within the protoplanetary disk and (2) the radial distribution of this mass determine the final multiplicity of planetary systems and the mutual orbital inclinations of the planets. In Section 2, we describe the details of our simulations, and in Section 3 describe their results. In Section 4, we compare observable quantities from our simulations: number of transiting planets, transit duration ratio, and period ratio, to \kepler\ observations. In Section 5, we describe the application of our simulations specifically to the sample of planets orbiting M dwarfs ($\S$5.1). We go on to comment upon the possible physical origin of the variation in the properties of planetesimal disks ($\S$5.2) and the long-term evolution of planetary systems ($\S$5.3). In $\S$5.4, we summarize the  predictions from our simulations and implications for future study, and conclude in Section 6.

%%%%%%%%%%%%%%%%%%%%%%%%%%%%%%%%%%%%%%%%%%%%%%%%%%%%%%%%%%%%%%
\section{Simulations}
The formation pathway of short period planets is a matter of debate. Models of their formation fall into two general categories: those where planets form in situ from a massive planetesimal disk \citep[e.g.][]{Hansen12, Chiang13} and those where planets form at larger distances from the star and later migrate to their currently observed locations \citep[e.g.][]{Terquem07, Cossou14}. Currently, comparison with known planetary systems does not definitively favor one model over the other. Although both models are able to form systems that generally match the architectures of observed planetary systems, neither of them have yet been able to duplicate all of their characteristics. It is expected that bodies larger than about the mass of Mars should feel significant torques from the disk \citep{Ward97, Tanaka02}; enough for substantial migration over the lifetime of a protoplanetary disk. On the other hand, a strong dependence on detailed physical structure of the disk leads to uncertainties in the magnitude and even direction of migration \citep{Laughlin04, Johnson06, Paardekooper06, Masset06, Kretke12, Terquem03} indicating we may not yet know enough about planet migration to definitively conclude that it is important for the formation of close-in planets. In this work, we assume that planets form in situ from a previously assembled planetesimal disk.

In order to evaluate this hypothesis, we ran a large set of N-body simulations of late-stage planet formation. We varied both the total mass of the planetesimal disk and the radial distribution of this mass. The initial surface density of the planetesimal disk is assumed to follow a power law:
\begin{equation}
	\Sigma = \Sigma_0 r^{\alpha}.
\end{equation}
We choose $\alpha$ to be -2.5, -1.5, -0.5 and set $\Sigma_0$ so that the total mass between 0.05 and 1 AU is 7, 10, 17 or 35 $M_{\oplus}$. For each combination of $\alpha$ and disk mass we ran 32 simulations with different realizations of the initial planetesimal disk.

Motivated by the results of \citet{Kokubo02}, we break the initial planetesimal disk into two populations: larger bodies, which we refer to as planetary embryos, and smaller bodies, which we refer to as planetesimals. Each type of body cumulatively accounts for 50\% of the total planetesimal disk mass. The initial masses and semi-major axes of the embryos are determined by enforcing two requirements: 1. their radial mass distribution follows the chosen surface density power law profile (divided by two since they only account for half the mass) and 2. they are separated by about 15 mutual hill radii. Depending on the surface density profile, this results in anywhere from 15 to 65 embryos with a range of masses (which scale with semi-major axis as described by Equation \ref{ExpectedMass}). For each embryo, we place ten planetesimals with a tenth of the embryo's mass randomly positioned half way between the orbits of the embryo's inner and outer-most nearest neighbors. The initial orbits of all bodies are close to circular ($e < 0.01$) and have low inclinations ($i < 1^\circ$). The remaining orbital elements are assigned random values between 0 and 360 degrees.

The evolution of each planetesimal disk around a solar mass star was integrated using the mercury N-body integrator \citep{Chambers99}. Collisions are assumed to result in perfect mergers of the bodies. The systems were integrated for 10 million years with a timestep of 0.5 days.

\section{Results}

We begin by constructing a simple model to predict the masses of planets as a function of semi-major axis given a set of initial disk conditions. In this model, the mass of a planet is determined by the size of its feeding zone and the amount of mass in that zone. We assume that the size of the feeding zone is similar to the distance between the planet and its neighbor and that this separation is some number, $\Delta$, times their mutual hill radius,
\begin{equation}
r_H \approx  \left(\frac{2M_{p}}{3M_*}\right)^{1/3}r_{p}.
\end{equation}
Given an initial power law surface density distribution and a separation between neighboring planets, $\Delta$, the expected mass of a planet forming at a given semi-major axis is:
\begin{equation}
M_{p}(r) = \frac{2\left(\Delta \pi \Sigma_0 r^{(\alpha+2)}\right)^{3/2}}{\sqrt{3M_*}}.
\label{ExpectedMass}
\end{equation}

\begin{figure*}     % use "figure*" instead of "figure" if you want your figure to span both columns
    % adjust this number to change the size of your figure
\includegraphics[width=7in9]{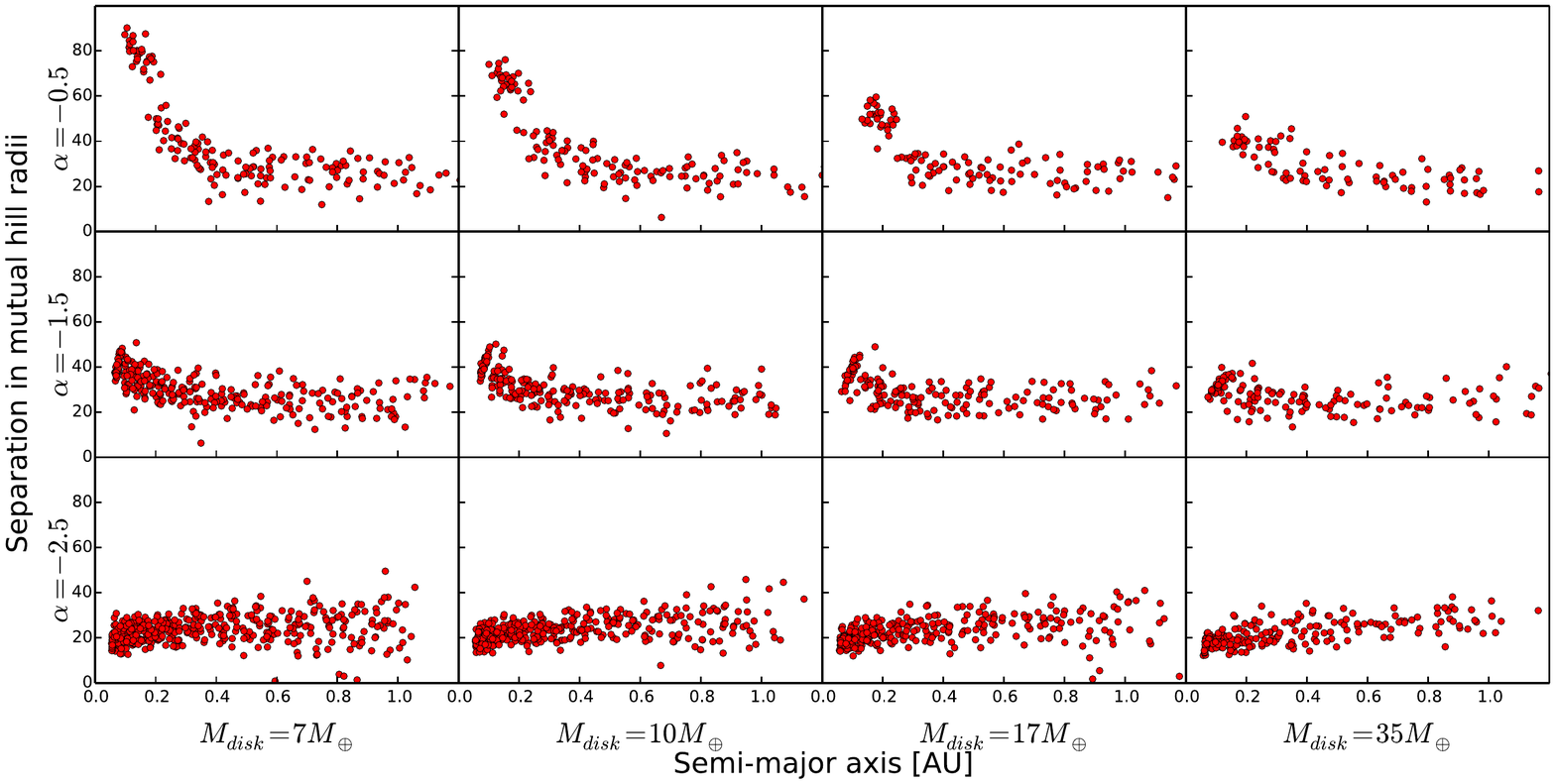}
\caption{Separation between adjacent planet pairs produced in simulations in units of mutual hill radii. Planets that form in disks with a shallower surface density profile tend to be more widely separated at smaller semi-major axes. Presumably this is due to their larger eccentricities. }
\label{HillSeparations}
\end{figure*}

Figure \ref{HillSeparations} shows the expected range of planetary masses (shaded in grey) assuming $15 < \Delta < 30$ (values typical for planetary systems formed in N-body simulations) for each different initial planetesimal surface density distribution compared to the actual mass of planets that formed in the simulations. For $\alpha=-2.5$, this simplistic model predicts, reasonably well, the masses of planets. For larger $\alpha$, the model is a poor predictor of planet mass. The failure of this model is not surprising considering that it ignores the complex dynamical interactions present in these systems. Here, we highlight two factors that influence the mass and spacing of planets: the radial redistribution of mass and the orbital eccentricity.

\begin{figure*}     % use "figure*" instead of "figure" if you want your figure to span both columns
%\epsscale{1.2}      % adjust this number to change the size of your figure
%\plotone{MassVsSMA.pdf}
\includegraphics[width=7in9]{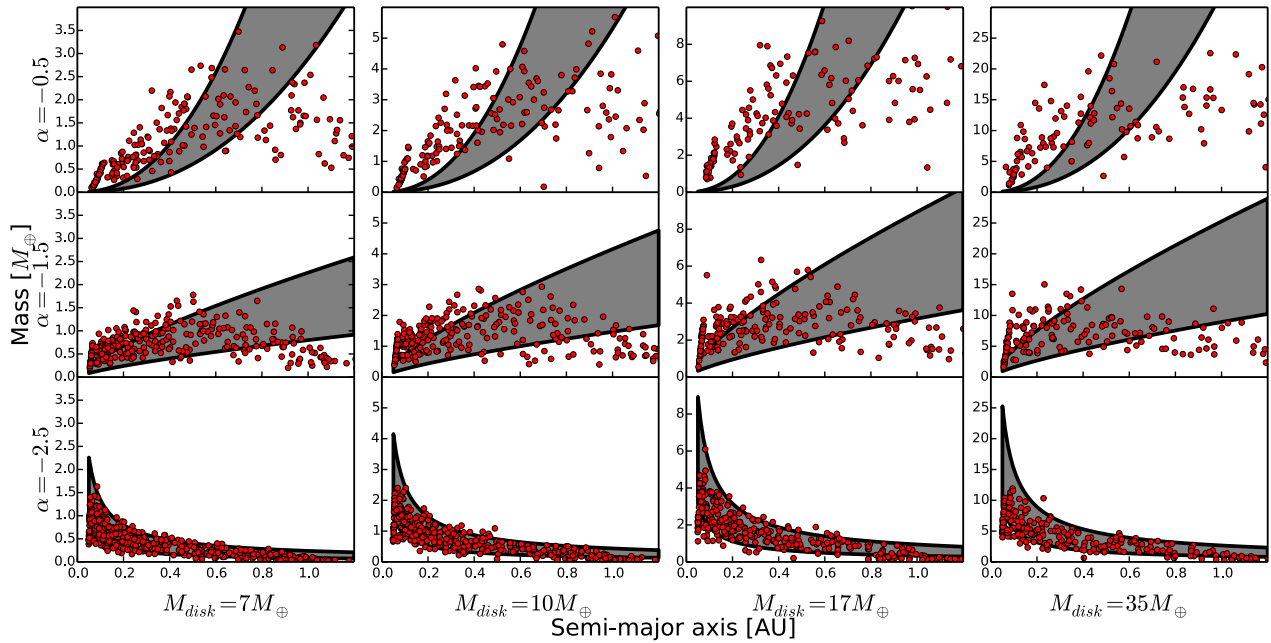}
\caption{Masses and semi-major axes of the planets produced in simulations. The grey shading indicates a first order approximation of their expected mass based on a naive model in which a planet accretes all material within its feeding zone. Deviations from this model are the result of the complex dynamics present in N-body simulations.}
\label{MassVsSma}
\end{figure*}

\begin{figure*}     % use "figure*" instead of "figure" if you want your figure to span both columns
    % adjust this number to change the size of your figure
\includegraphics[width=7in9]{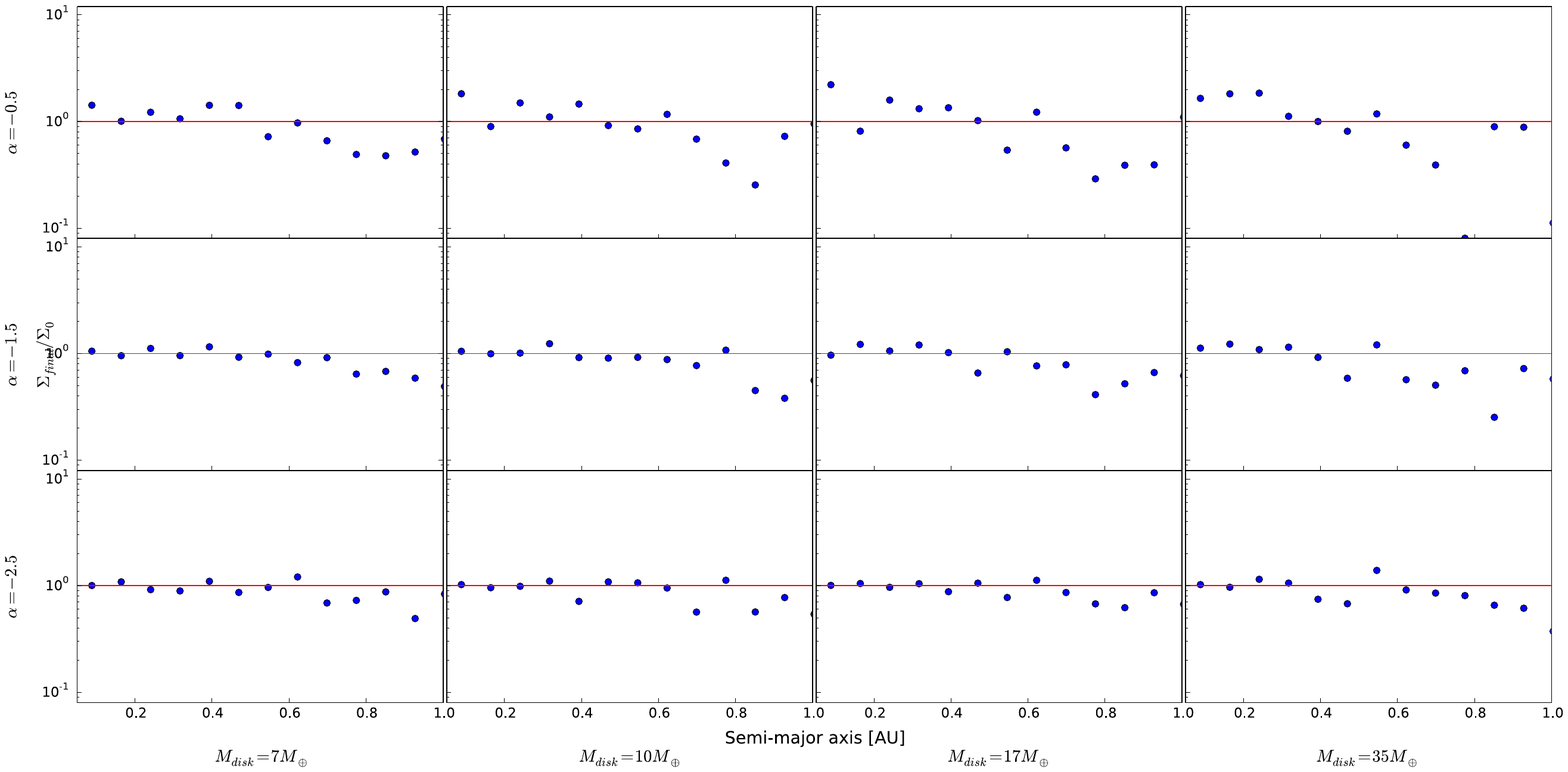}
\caption{The average final surface density of simulated planetary systems divided by the initial surface density of the planetesimal disk as a function of semi-major axis. The final systems tend to be under-massive beyond about 0.6 AU compared to the initial planetesimal disk as a consequence of the redistribution of mass from planetesimal-planetesimal scattering.}
\label{Redistribution}
\end{figure*}

Equation \ref{ExpectedMass} is only valid if the final mass surface density distribution is the same as the initial one used to derive the equation. However, the numerous gravitational scatterings and resonant interactions that occur in the N-body simulations can significantly redistribute mass in the system. Figure \ref{Redistribution} shows the averaged surface density distributions of the final planetary systems relative to the initial planetesimal surface density distributions. It is evident that there is some amount of mass redistribution in all systems. In particular, the outer disk, beyond $\sim$0.6 AU loses mass during the simulation. Some of this mass is scattered outward and to conserve energy some must move inward. The inward movement of material is more apparent in the shallower surface density profile disks because there is less mass in the inner disk to begin with and any addition to this is more noticeable. This decrease of mass in the outer disk is reflected in Figure \ref{MassVsSma} in which the simulated planets are undermassive compared to what is expected from Equation \ref{ExpectedMass}.

\begin{figure*}     % use "figure*" instead of "figure" if you want your figure to span both columns
    % adjust this number to change the size of your figure
\includegraphics[width=7in9]{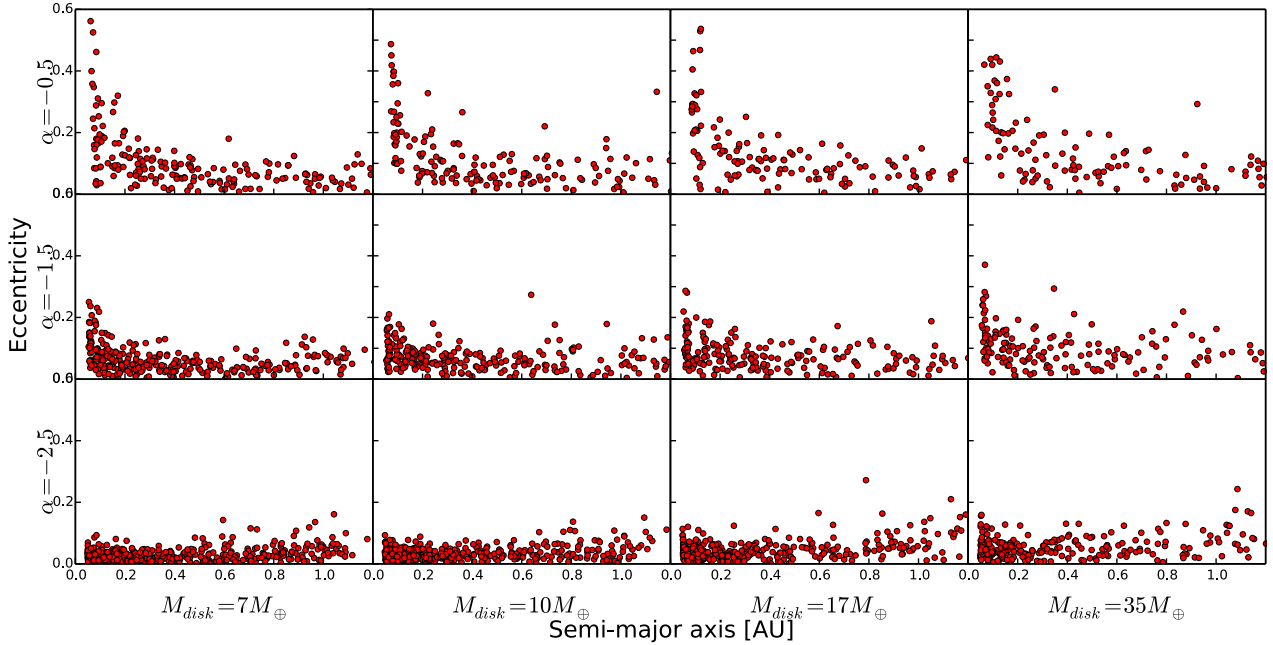}
\caption{Orbital eccentricities as a function of semi-major axis for planets produced in the simulations. There is a significant rise in eccentricity at smaller semi-major axes and an overall increase in eccentricity for planets that form from planetesimal disks with shallower surface density profiles. }
\label{EccentricityVsSma}
\end{figure*}

The eccentricity affects the final mass of a planet because it changes the width of that planet's feeding zone. A planet on an eccentric orbit will occupy a larger radial range of the disk than one on a circular orbit. This effectively increases the value of $\Delta$ in Equation \ref{ExpectedMass} leading to a larger expected mass. This affect can be seen in Figures \ref{HillSeparations} and \ref{EccentricityVsSma}. For $\alpha$ = -1.5 and -0.5, in the inner disk there is a rise in eccentricity and a corresponding rise in the orbital separations between neighboring planets. Thus we would expect for this region that the masses of planets that form in the simulations are larger than our simple prediction from Equation \ref{ExpectedMass}. Indeed, this is what we see in Figure \ref{MassVsSma}.

\begin{figure*}     % use "figure*" instead of "figure" if you want your figure to span both columns
    % adjust this number to change the size of your figure
\includegraphics[width=7in9]{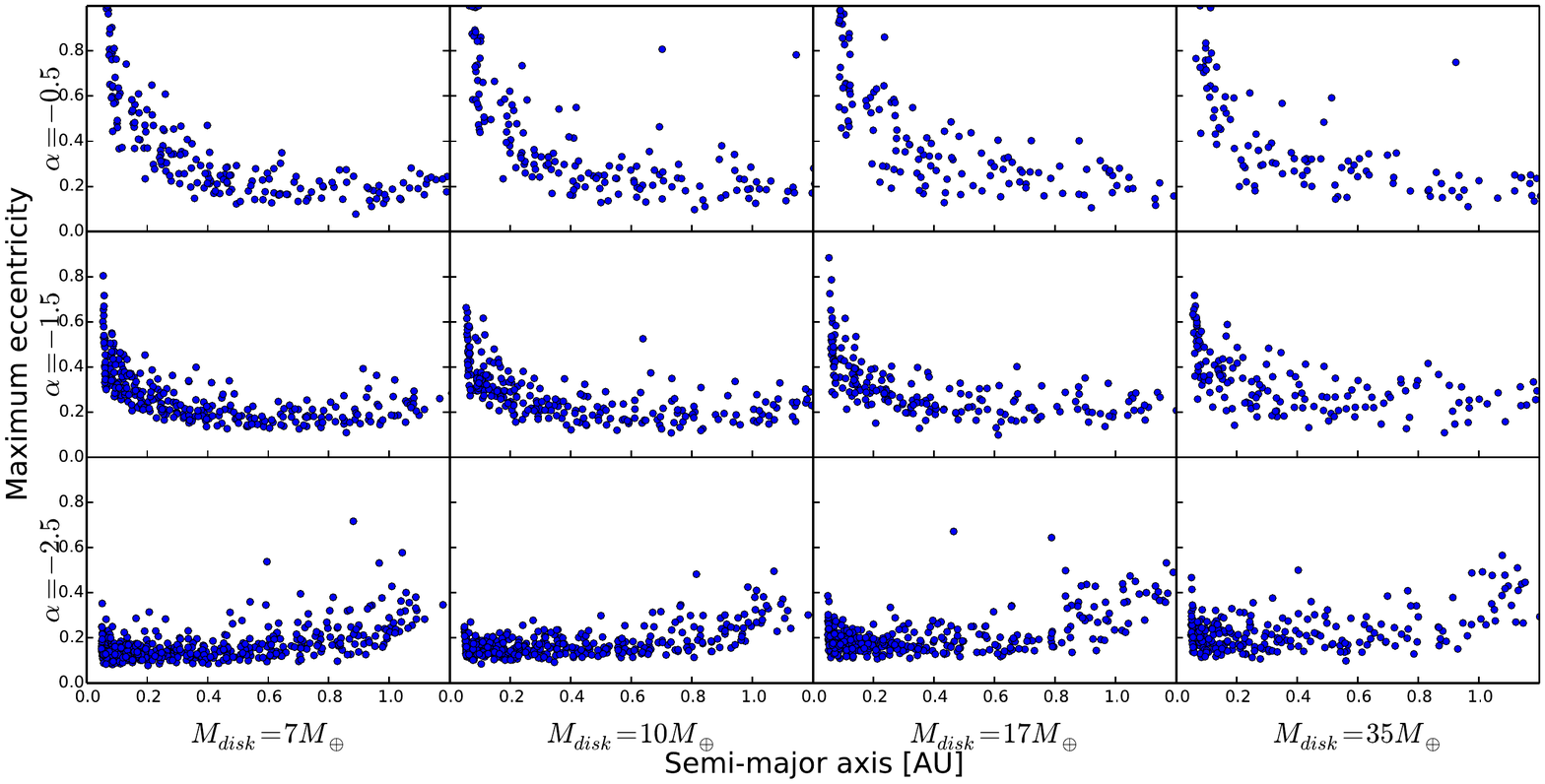}
\caption{The maximum eccentricity of each simulated planet assuming all the angular momentum deficit of that system is deposited on that planet. The same general trend of eccentricity with semi-major axis is seen in the maximum eccentricities as in the actual eccentricities suggesting that secular chaos plays a large part in determining the eccentricities of the planets formed in these simulations. }

\label{maxeccentricity}
\end{figure*}

We can start to understand the trends in eccentricity and planet-planet separations with semi-major axis by considering the angular momentum of these systems, or alternatively, the angular momentum deficit (AMD). The AMD is the difference in angular momentum that a planetary system has compared to what it would have if the orbits were circular and coplanar:

\begin{equation}
AMD=\sum_{k=1}^N \Lambda_k \left(1-\sqrt{1-e_k^2}cos(i_k)\right),
\label{amd}
\end{equation}
where $\Lambda_k = \frac{m_k M_{\odot}}{m_k+M_{\odot}}\sqrt{G(m_k+M_{\odot})a_k}$ is the circular angular momentum of the $k$th planet. The AMD of a system provides some measure of its stability. A system with zero AMD (composed of coplanar and circular orbits) will remain stable forever, whereas a system with sufficiently high AMD will exhibit secular chaos \citep{Wu11}. Secular chaos enables the transfer of AMD between planets leading to potentially large values of their eccentricity and inclination. Because AMD is conserved in secularly interacting systems, the maximum eccentricity a planet can attain can be calculated from Equation \ref{amd} by assuming the planet accounts for the entirety of the AMD for the system (i.e. the other planets are on circular and coplanar orbits). This is not the case for systems with resonant interactions, as in many of our simulations. However the AMD varies by no more than a factor of a few after the first million years for most of our simulations and so the maximum eccentricity calculation is still approximately true. Figure \ref{maxeccentricity} shows the maximum eccentricity calculated for each of our simulated planets given the AMD, masses and semi-major axes of the planets at the end of the simulation. The same general trends of eccentricity with semi-major axis and initial planetesimal disk structure are apparent in the calculated maximum eccentricities and the actual eccentricities albeit with an offset in eccentricity, indicating that secular chaos is a likely cause for the trends in eccentricity. For the simulations with $\alpha = -0.5, -1.5$ planet mass increases with semi-major axis. Accordingly, the angular momentum of the inner planets is much smaller than that of the outer planets. In this case, a moderate fraction of the total AMD of the system can be large compared to the angular momentum of one of the inner planets. Thus if this AMD is transferred to that planet, its eccentricity can become quite large. On the other hand, in the simulations with $\alpha=-2.5$, masses decrease with semi-major axis. The angular momentum of individual planets in these systems are more similar and accordingly their maximum eccentricities do not vary significantly with semi-major axis.

\begin{figure*}     % use "figure*" instead of "figure" if you want your figure to span both columns
    % adjust this number to change the size of your figure
\includegraphics[width=7in9]{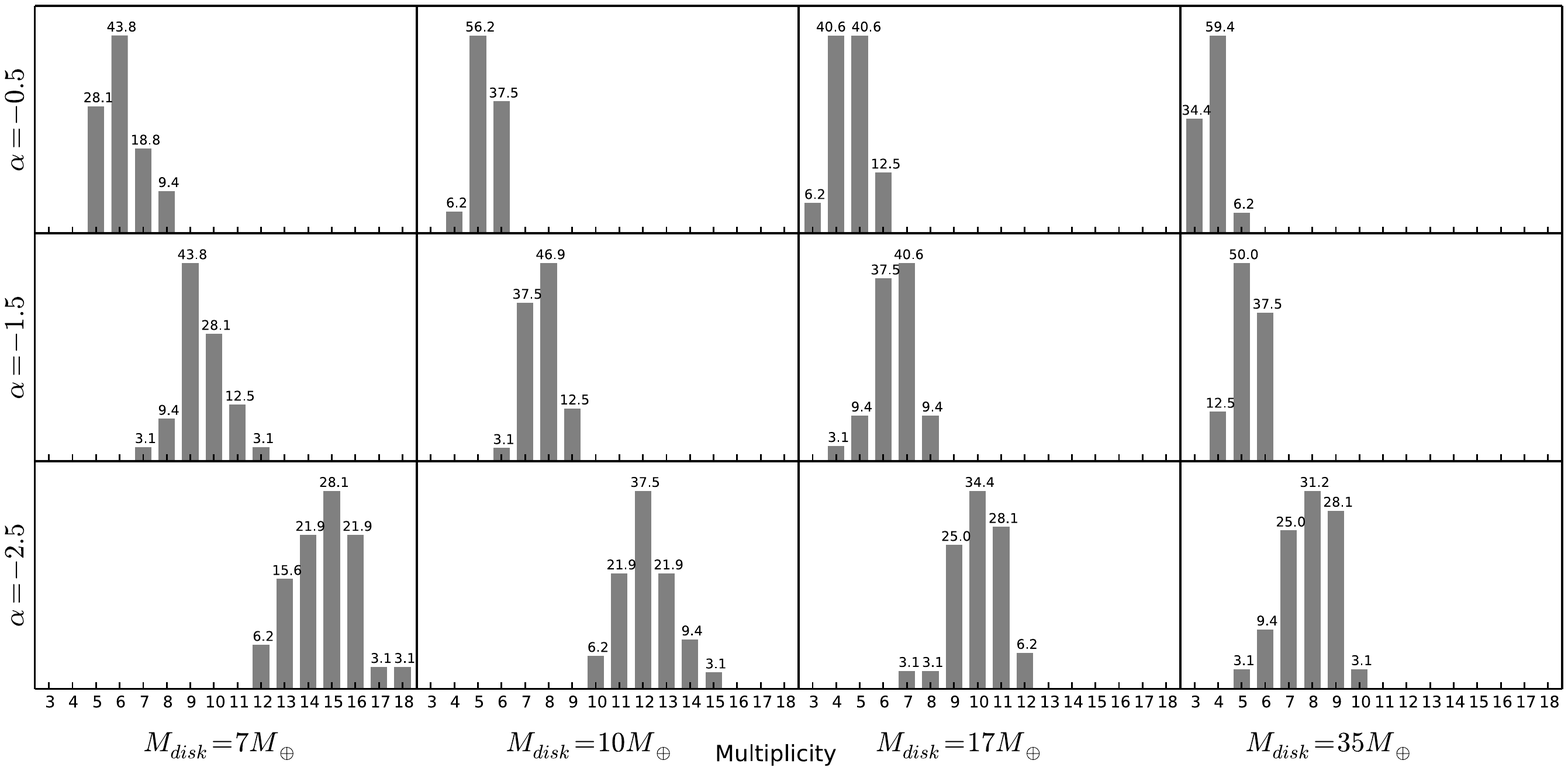}
\caption{Multiplicity of systems formed with different initial planetesimal disk conditions. This includes all bodies remaining at the end of each simulation. }
\label{IntrinsicMultiplicity}
\end{figure*}

The distribution of masses and orbital separations of planets ultimately determine the final multiplicity of the planetary systems. Given some mass budget, an increase in orbital separation and the corresponding increase in planet mass will result in fewer planets in the system. Because the systems formed in simulations with larger values of $\alpha$ form, on average, more massive and more widely separated planets, we expect their multiplicity to be lower. This is demonstrated in Figure \ref{IntrinsicMultiplicity}. Changing the initial distribution of mass in the planetesimal disk leads to a significant change in the multiplicity of systems that form. Larger $\alpha$ and higher disk mass correspond to the formation of significantly fewer planets. As we will show in Section \ref{comparison}, this has a strong impact on the observed multiplicity of planetary systems.

\begin{figure*}     % use "figure*" instead of "figure" if you want your figure to span both columns
    % adjust this number to change the size of your figure
\includegraphics[width=7in9]{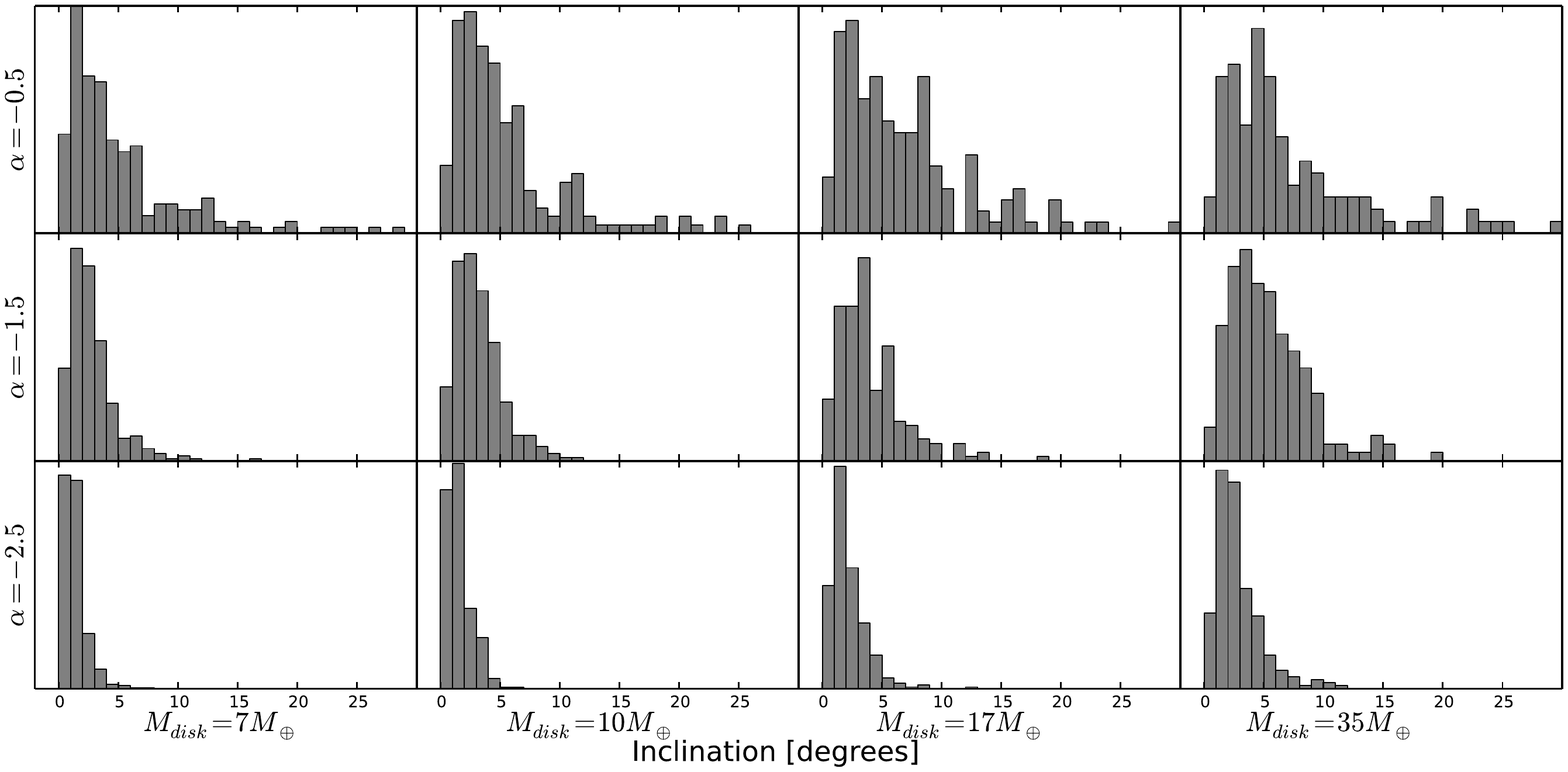}
\caption{Distribution of orbital inclinations of planets formed in simulations with different initial planetesimal disk conditions. }
\label{Inclinations}
\end{figure*}

The other determining factor in the observed multiplicity of systems is the spread in mutual inclinations between planets. The effects of the initial planetesimal disk conditions on final planet inclination are significant. More massive disks tend to produce planets with moderately higher mutual inclinations and disks with larger $\alpha$ produce planets with significantly larger mutual inclinations (see Figure \ref{Inclinations}).

\section{Comparison to Observations}
\label{comparison}

In order to compare our simulated planetary systems to the catalog of known exoplanets, we ``observe" the final state of each dynamical simulation 10,000 times. We simulate the viewing geometry isotropically. Our final synthetic yield of transiting planets are those (1) with impact parameter $b<1$ (from the given viewing angle) and (2) whose probability of detection by \kepler\ exceeds a randomly drawn uniform number between 0 and 1.

We first compare our simulations to observations of {\it Kepler} transiting planets orbiting GK stars. To create a realistic synthetic sample to compare to the observed sample of transiting planets, we employ the \kepler\ pipeline completeness calculation. \citet{Burke15} contains a likelihood of detection as a function of orbital period and planetary radius. This model uses the completeness parameterization presented in \citet{Christiansen15}, itself generated  from signal injection and recovery in \kepler\ data. Because our simulations calculate planet mass, we translate the planetary masses to radii to determine a detection probability for each planet. We use the probabilistic relationship presented in \citet{Wolfgang15b} for $M_p > 2.13 M_{\oplus}$ (corresponding to $R_p > 1.2 R_{\oplus}$, with the restriction that the radius must be greater than 1.2 $R_{\oplus}$. For planets with masses less than $2.13 M_{\oplus}$, we use the best fit deterministic model from \citet{Wolfgang15b} because the data does not rule out a deterministic mass/radius relation for small planets and the scatter in the probabilistic model is unphysical for small planets.
Our comparison sample consists of all known planets orbiting the stars used to calculate the detection probability grid. This sample is made up of 2,653 planets in 2042 systems.

\subsection{Multiplicity Distribution}

\begin{figure*}     % use "figure*" instead of "figure" if you want your figure to span both columns
    % adjust this number to change the size of your figure
\includegraphics[width=7in9]{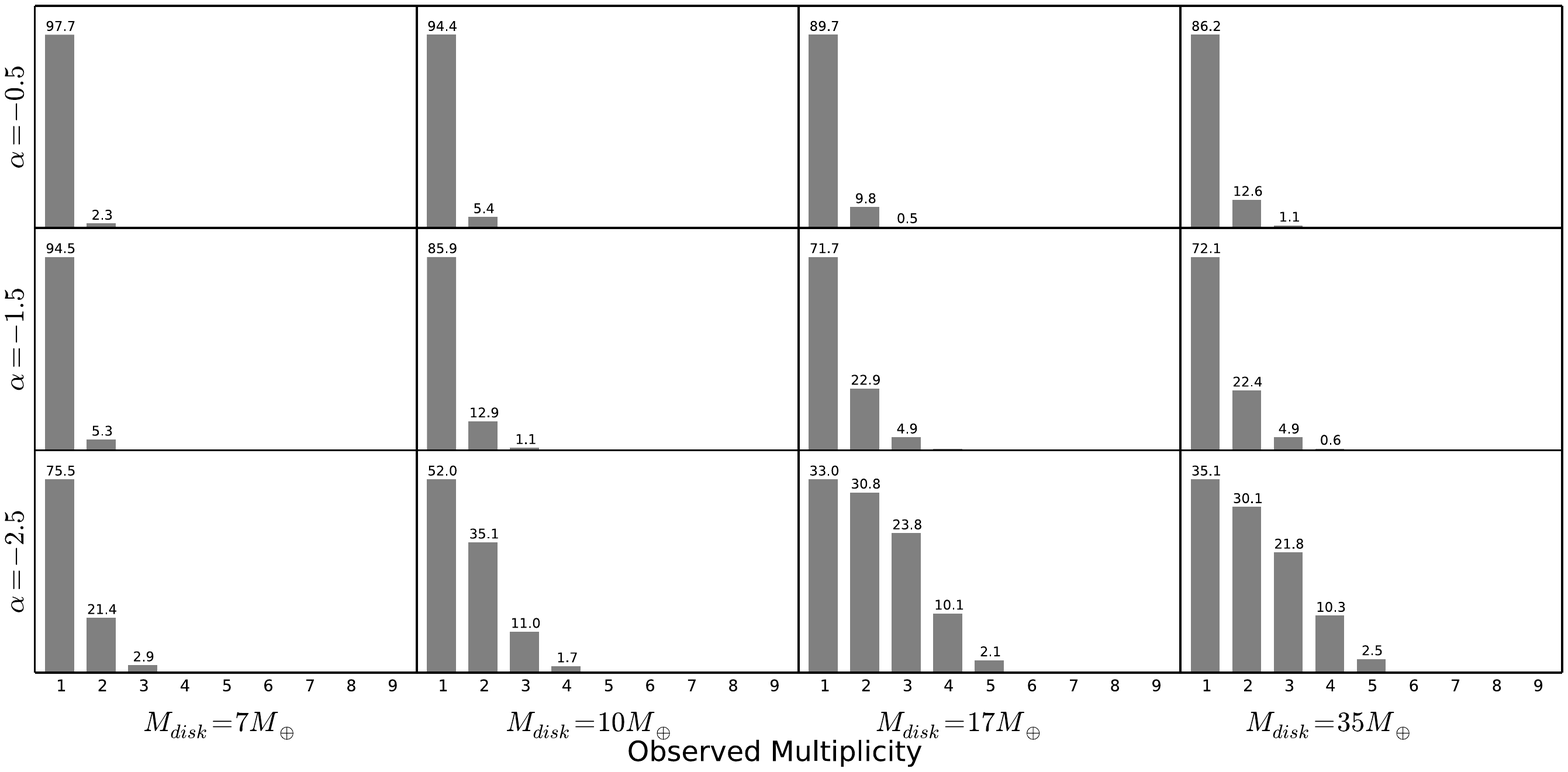}
\caption{The multiplicity distribution of the simulated systems that would be observed by \kepler.}
\label{ObservedMultiplicity}
\end{figure*}

Figure \ref{ObservedMultiplicity} shows the distribution of apparent multiplicities from our set of ``observed" systems for each initial disk structure. There is a clear dependence of the observed multiplicity on the initial planetesimal disk distribution, showing variations with both the surface density power law index, $\alpha$, and the total initial mass in the disk, $M_{\mbox{tot}}$.

Our observed data set $\{D\}$ is a vector with six entries, corresponding to the number of stars hosting $n$ tranets, where 1$<n<$6. In our final \kepler\ sample, there are 1635 G and K dwarfs hosting 1 detected transiting planet, 269 hosting 2, 88 hosting 3, 36 hosting 4, 12 hosting 5, and 2 hosting 6.

%We compare the outcome of our simulations to the observed \kepler\ tranets as follows. We consider only planets with orbital periods $>$ 4 days, per the innermost orbital periods included in our dynamical simulations. We note that including planets with orbital periods interior to 4 days increases the sample only by 10\%. Our final observed data set $\{D\}$ is a vector with six entries, corresponding to the number of stars hosting $n$ tranets, where 1$<n<$6. In our final \kepler\ sample, there are 401 G and K dwarfs hosting 1 detected transiting planet, 70 hosting 2, 28 hosting 3, 9 hosting 4, 1 hosting 5, and 1 hosting 6.

We compare the model--predicted population, $\mu_{n}$ (with results shown in Figure \ref{ObservedMultiplicity}) to the observed yield $D_n$ of tranets. We estimate the likelihood of the model we have generated from our simulations, $\mu_{n}$, given $D_n$ assuming the number of systems with $n$ transiting planets is described by Poisson counting statistics. Therefore, we can describe the likelihood of the model as:

%Poisson counting statistics describe integer numbers of transiting planets, so we evaluate the likelihood of $\alpha$ and $M_{\mbox{tot}}$ with a Poisson likelihood function, which is conditioned on the observed number of stars hosting $n$ planets, $D_n$, with the ensemble of bins given by $\{D\}$. Therefore, we can describe the likelihood $\mathcal{L} \equiv P(\{M\}|N,\sigma)$ of observing the distribution $\{D\}$:

\begin{equation}
\mathcal{L} \propto \prod_{n}\frac{\mu_{n}^{D_n}e^{-\mu_{n}}}{D_n!},
\label{eq:bayes1}
\end{equation}
where $\mu \equiv \mu(\alpha, M_{\mbox{tot}})$ from our dynamical simulations, and $n$ runs from 1 to 6.

We first evaluate $\mathcal{L}$ using only one mode of planet formation: that is, assuming one value for $\alpha$ and $M_{\mbox{tot}}$. Our grid of $\alpha$ and $M_{\mbox{tot}}$ consists of 12 points, with 3 possible values of $\alpha$ (-0.5, -1.5, and -2.5) and 4 possible values of $M_{\mbox{tot}}$ (7, 10, 17, 35 $M_{\oplus}$). The model distribution, $\mu_{n}$, is created by re-normalizing the simulated multiplicity distribution (shown in \ref{ObservedMultiplicity}) to the number of systems in our observed sample. We also employ a mixture model, to test whether the data support a model with a range of initial conditions for planet formation. We assume the simplest case of two modes, each with independent $\alpha$ and $M_{\mbox{tot}}$, occurring with respective frequency $f$ and $(1-f)$. The model distribution, $\mu_{n}$, is the weighted sum of the multiplicity distributions of each mode. We assign a flat prior on both $\alpha$ and $M_{\mbox{tot}}$, and employ the Bayesian sampler MultiNest \citep{Feroz08, Feroz09, Feroz13} to evaluate the posterior distributions.

%We also employ a mixture model, to test whether the data support the hypothesis of multiple distinct modes of planet formation. We assume the simplest case of two modes, each with independent $\alpha$ and $M_{\mbox{tot}}$, occurring with respective frequency $f$ and $(1-f)$.

We compare the output number of tranets from our simulations to the observed tranets of the \kepler\ GK host stars. In agreement with previous studies, we find that the data favor a two-component model over a single component model.  Taking the ratio of the Bayesian evidences returned by Multinest, we find a log evidence for a mixture model of -8.8, and log evidence of -26.7 for a single mode model. We conclude that the data are better supported by a mixture model by a factor of 18:1.

%For G dwarfs:
%Mixture model log(evidence)= -8.8
%ingle mode log(evidence)= -26.7

%For M dwarfs:
%Mixture model log(evidence): -8.4
%Single mode log(evidence) = -10.5

\begin{figure*}     % use "figure*" instead of "figure" if you want your figure to span both columns
      % adjust this number to change the size of your figure
\includegraphics[width=7in]{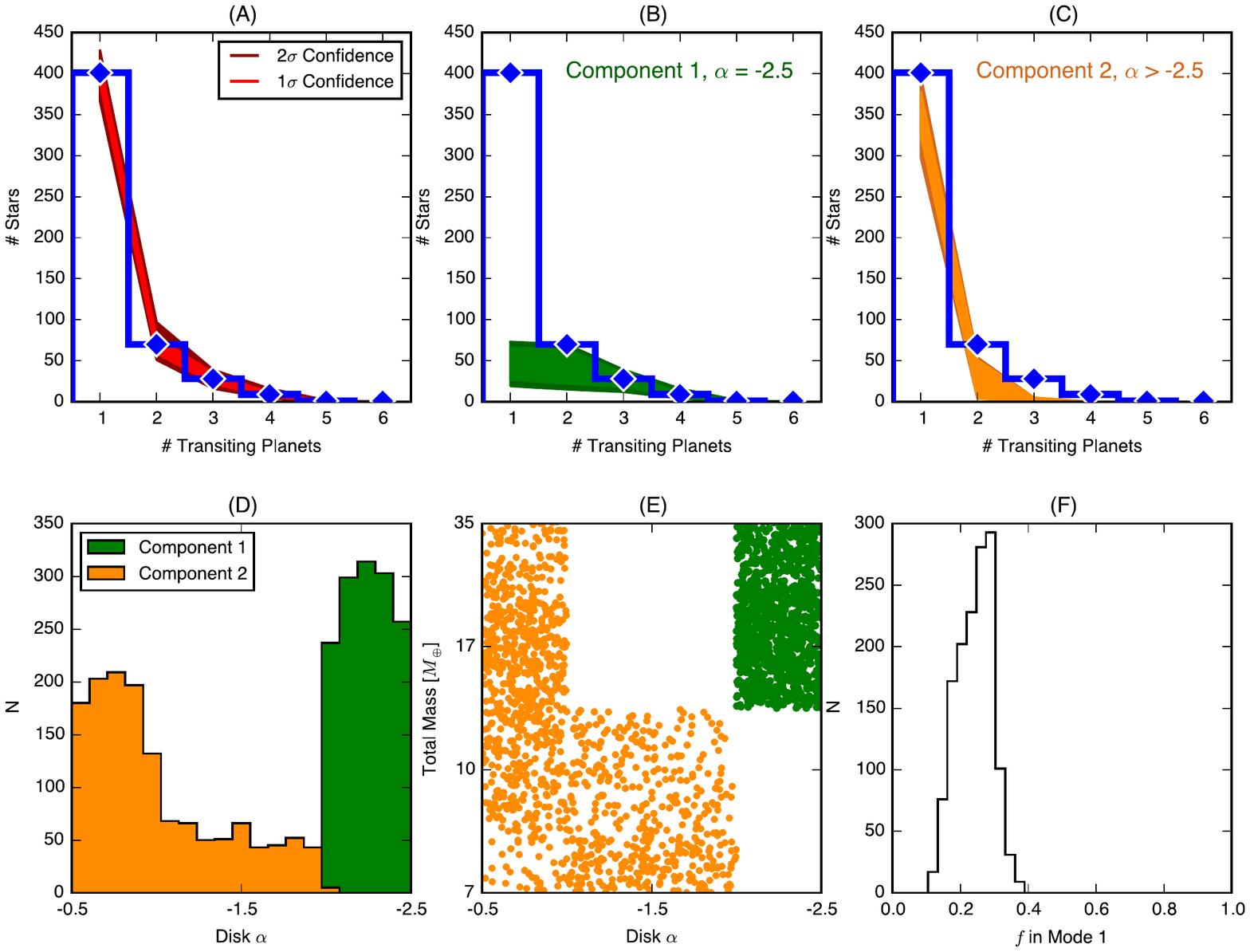}
\caption{Clockwise from top left: (A) Observed distribution of tranets in blue, with 1 and 2$\sigma$ confidence model region overplotted in light and dark red, respectively. (B) 1$\sigma$ region of first component to mixture model. (C) 1$\sigma$ region of second component to mixture model. (D) Location of MCMC realizations in disk $\alpha$ and $M_{\mbox{tot}}$, in green for $\{ \alpha_{1},M_{\mbox{tot},1} \}$ and orange for $\{ \alpha_{2},M_{\mbox{tot},2} \}$. (E) Marginalized posterior in $\alpha_{1}$ and $\alpha_{2}$. (F) Marginalized posterior in $f$, fraction of mixture model in first component.}
\label{fig:tranets_fit}
\end{figure*}

In Figure \ref{fig:tranets_fit}, we see that the high multiplicity tail of the multiplicity distribution requires one of the two modes to correspond to systems formed in disks with the steepest surface density profile ($\alpha$=-2.5) and a disk mass of 17 Earth-masses or greater. The mode that accounts for the high number of singly transiting systems is less well constrained because many of the disk configurations lead to systems that typically show only one transiting planet. However, models with shallower surface density profiles are favored. We find that between 15 and 35\% of systems must form in the first mode while the rest form in the second. This finding resembles that of \citet{Fang12b} who find that 75-80\% of systems account for the majority of low observed multiplicity systems. However, our conclusions regarding the underlying architecture of these systems differ. They find that most systems are made up of 1-2 planets with low mutual inclinations, whereas we find the underlying multiplicity and mutual inclinations to be substantially larger.

\subsection{Duration Ratio and Period Ratio Prediction}

Studies such as \cite{Hansen13}, \cite{Fang12b}, and \cite{Fabrycky14} employ ensemble normalized transit durations in multi-tranet systems to infer average orbital mutual inclination.

Transit duration depends upon orbital period as well as impact parameter. The ratio of transit durations, $\xi$ (normalized by orbital velocity, per Equation \ref{eq:duration}), of adjacent planets therefore encodes information about their mutual inclination.

\begin{equation}
\xi = \frac{\tau_{\mbox{inner}} / P_{\mbox{inner}}^{1/3}}{\tau_{\mbox{outer}}/ P_{\mbox{outer}}^{1/3}}
\label{eq:duration}
\end{equation}

\begin{figure*}     % use "figure*" instead of "figure" if you want your figure to span both columns
      % adjust this number to change the size of your figure
\includegraphics[width=7in]{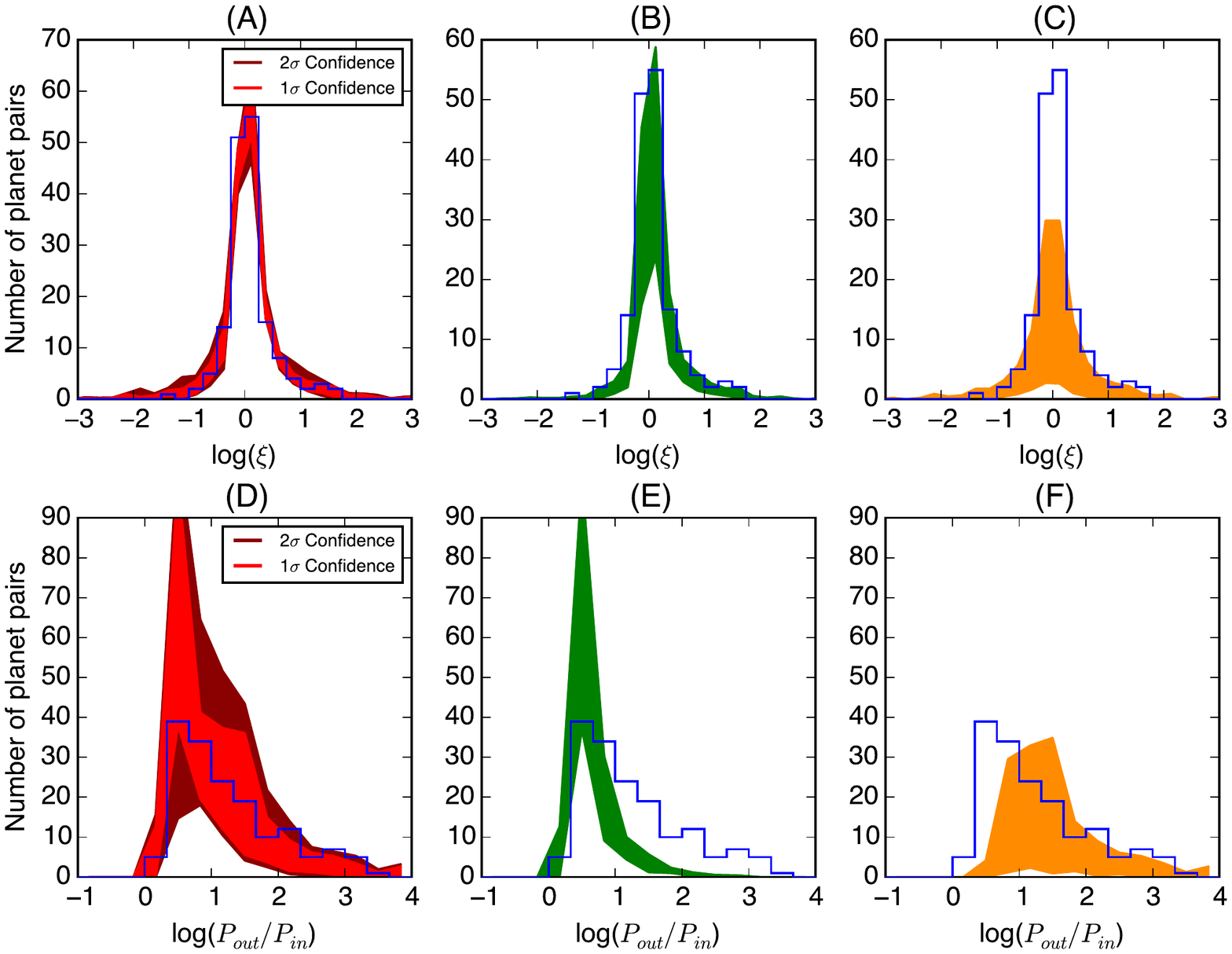}
\caption{Predictions for period and duration ratio distributions, inferred from tranet fits for GK dwarfs. Clockwise from top left: (A) Observed distribution of log of duration ratio, log($\xi$), in blue. 1 and 2$\sigma$ confidence model region are overplotted in light and dark red, respectively. (B) 1$\sigma$ region of first component to mixture model. (C) 1$\sigma$ region of second component to mixture model.(D) Observed distribution of log($P_{\mbox{out}}/P_{\mbox{in}}$) in blue, with 1 and 2$\sigma$ confidence model region overplotted in light and dark red, respectively. (E) 1$\sigma$ region of first component to mixture model. (F) 1$\sigma$ region of second component to mixture model.}
\label{fig:perratio_fit}
\end{figure*}

In Figure \ref{fig:perratio_fit}, we show the \kepler\ observed distribution of orbit-normalized transit durations and the \kepler\ observed distribution of period ratios for transiting planets (both in blue). We overplot our model {\it predictions} for these distributions that we infer from our multiplicity fit. We see that the two-component model self-consistently recovers not only the tranet distribution, but also the duration and period ratio distributions (neither of which are well-approximated by a single model).

\section{Discussion}

\subsection{Application to M dwarf Systems}
We also examine the effect of stellar mass upon our simulations. There exists suggestive evidence that planet formation proceeds differently around M dwarfs than around GK dwarfs. The relationship between stellar metallicity and planet occurrence may vary as a function of stellar mass \citep{Gaidos14}, as well as the mass distribution of the planets themselves \citep{Howard12, Mulders15}. M dwarfs host, on average, 2.5 planets per star, many more than sun-like stars \citep{Dressing15b, Morton14b}. Intriguingly, both GK dwarfs \cite{Lissauer11} and M dwarfs \cite{Ballard14} also exhibit a dichotomy in their underlying architecture. For M dwarfs, a mixture model best recovers the observed multiplicity distribution. Half of these systems contain at least 5 planets with low mutual inclinations, and the other half containing either one planet or multiple planets with high mutual inclinations. The difference in multiplicity distribution between GK dwarfs and M dwarfs is also suggestive: 1/3rd of M dwarfs planet hosts uncovered by \kepler\ host 2 or more transiting planets, while this figure is 1 in 5 for GK dwarfs. However, \kepler's sensitivity to transits varies with stellar mass. A meaningful comparison requires folding \kepler's completeness to planets into our multiplicity study.

To address this, we performed a separate set of simulations with a stellar mass of 0.5 $M_{\odot}$. This value corresponds to the mass of the typical M dwarf observed by \kepler. For this set of simulations, we began with planetesimal disks extending from 0.05 - 0.5 AU. Rather than varying total mass in the disk as with our GK dwarf simulations, we make the simplifying assumption that the disk contains 10 $M_{\oplus}$ of material. We varied the surface density profile by changing the power law index to be -2.5, -1.5, -0.5, and 0.0. For each value of the power law index we ran eight simulations.

\begin{figure*}     % use "figure*" instead of "figure" if you want your figure to span both columns
      % adjust this number to change the size of your figure
\includegraphics[width=7in]{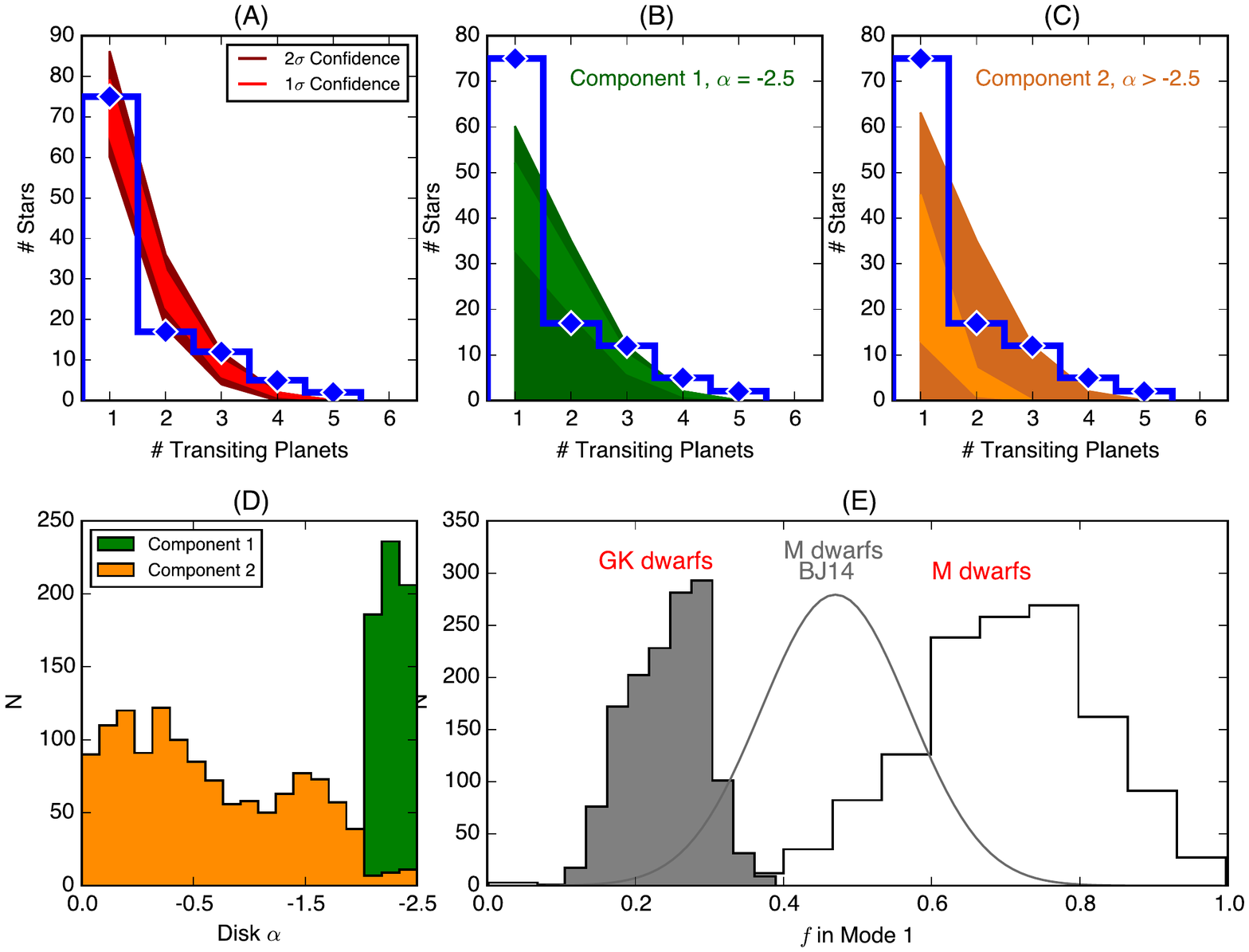}
\caption{Clockwise from top left: (A) Observed distribution of tranets in M dwarf systems in blue. 1 and 2$\sigma$ confidence model region is overplotted in light and dark red, respectively. (B) 1$\sigma$ region of first component to mixture model. (C) 1$\sigma$ region of second component to mixture model. (D) Location of MCMC realizations in disk $\alpha$ and $M_{\mbox{tot}}$, in green for $\{ \alpha_{1},M_{\mbox{tot},1} \}$ and orange for $\{ \alpha_{2},M_{\mbox{tot},2} \}$. (E) Marginalized posterior in $\alpha_{1}$ and $\alpha_{2}$. (F) Marginalized posterior in $f$, fraction of mixture model in first component. We have overplotted the distribution in $f$ for GK dwarfs from Figure \ref{fig:tranets_fit}, as well as the posterior from \cite{Ballard14}.}
\label{fig:tranets_mstars_fit}
\end{figure*}

For comparison with the \kepler\ sample, we generated synthetic ``observations" in the same way we did for the solar-mass sample. We employed the detection efficiency grid for planets around M dwarfs presented in \citet{Dressing15b}. We used the sample of planets detected in that work to ensure the detection efficiency was the same for the observed sample and our simulated sample. The results of the tranet fit are shown in Figure \ref{fig:tranets_mstars_fit}.

We find that the data favor a mixture model by a more modest value for M dwarfs than for the GK dwarfs. A mixture model in this case returns a log evidence of -8.4, while a single mode model returns -10.5. The 2:1 evidence ratio we find in this work is less than 8:1 stated by \cite{Ballard14}. We attribute this difference to the relative poorness-of-fit of even the mixture model in this case. Because we limited our suite of M dwarf simulations to those with a total mass of 10$M_{\oplus}$, we cannot well reproduce the high tranet multiplicity that we observe. Within our GK simulations, we found that we required total masses of 17 $M_{\oplus}$ to recover the shape of the multiplicity distribution. We infer that increasing the available mass in our set of M dwarf simulations would similarly result in a better fit to observations.

\begin{figure*}     % use "figure*" instead of "figure" if you want your figure to span both columns
      % adjust this number to change the size of your figure
\includegraphics[width=7in]{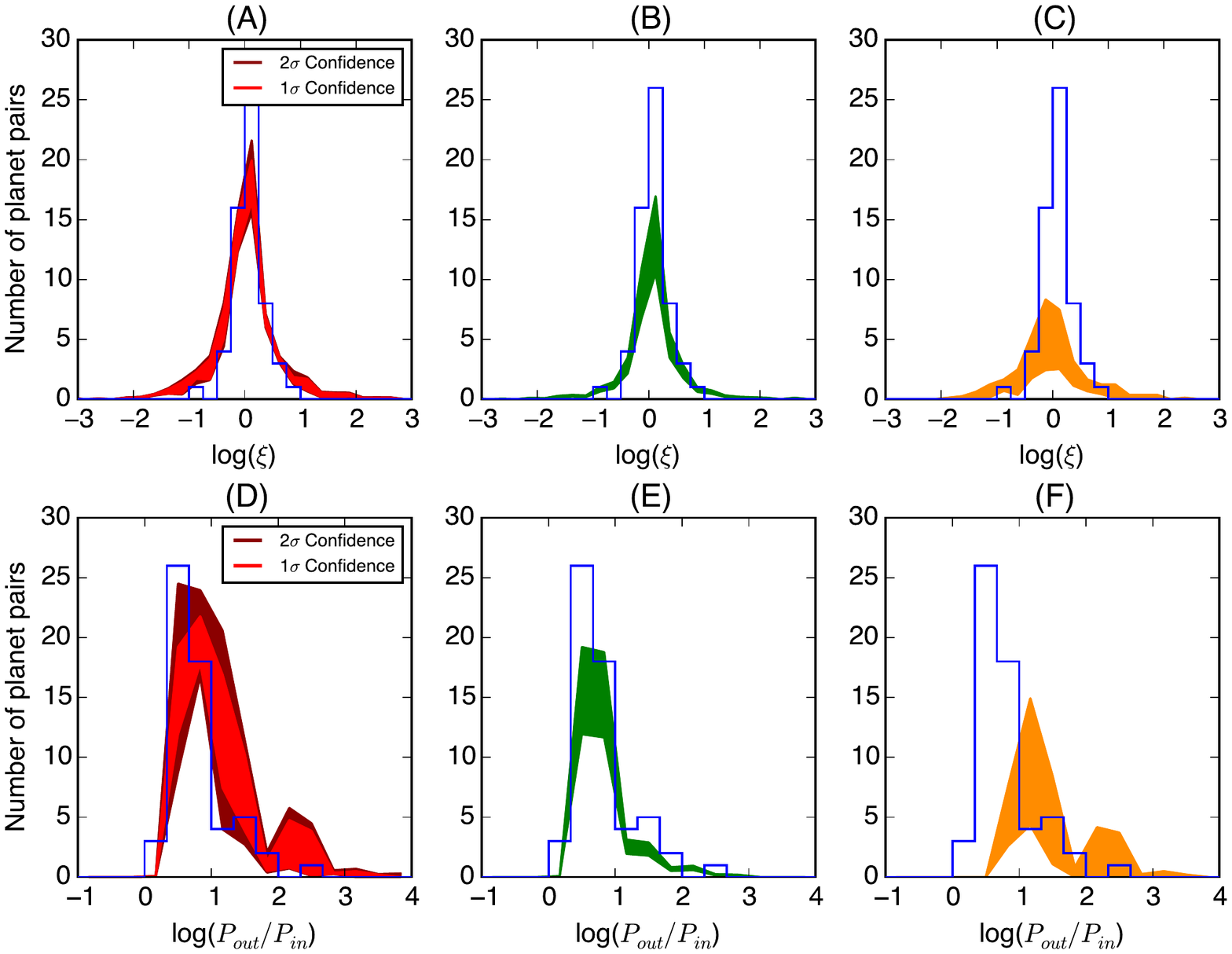}
\caption{Predictions for period and duration ratio distributions, inferred from tranet fits for M dwarfs. Clockwise from top left: (A) Observed distribution of log of duration ratio, log($\xi$), in blue. 1 and 2$\sigma$ confidence model region are overplotted in light and dark red, respectively. (B) 1$\sigma$ region of first component to mixture model. (C) 1$\sigma$ region of second component to mixture model.(D) Observed distribution of log($P_{\mbox{out}}/P_{\mbox{in}}$) in blue, with 1 and 2$\sigma$ confidence model region overplotted in light and dark red, respectively. (E) 1$\sigma$ region of first component to mixture model. (F) 1$\sigma$ region of second component to mixture model.}
\label{fig:mstars_plots}
\end{figure*}

%\begin{figure*}     % use "figure*" instead of "figure" if you want %your figure to span both columns
%      % adjust this number to change the size of your figure
%\includegraphics[width=7in]{disk_multinest_mstars_posterior_hill.pdf}
%\caption{Predictions for period and duration ratio distributions, inferred from tranet fits for GK dwarfs (top row) and M dwarfs (bottom row). Contributions from mixture model indicated in orange (steeper surface density law) and green (shallower surface density power law).}
%\label{fig:hill_separation}
%\end{figure*}

%For our M dwarf fit, the fit improvement we find with a mixture model is more modest than for the GK dwarfs. A mixture model in this case returns a log evidence of -8.4, while a single mode model returns -10.5. The 2:1 evidence ratio we find in this work is less than 8:1 stated by \cite{Ballard14}. We attribute this difference to the relative poorness-of-fit of even the mixture model in this case. Because we limited our suite of M dwarf simulations to those with a total mass of 10$M_{\oplus}$, we cannot well reproduce the high tranet multiplicity that we observe. Within our GK simulations, we found that we required total masses of 17 $M_{\oplus}$ to recover the shape of the multiplicity distribution.

The fractions $f$ of M dwarf planetary systems that reside in the flatter geometry is 63$\pm$16\%, versus the 24$\pm$7\% of GK planetary systems. This value is consistent with the value of $f$  from \cite{Ballard14} for M dwarf planetary systems of 47$\pm$10\%. The hypothesis of two distinct values for $f$ is favored by 611:1, as compared to the hypothesis that these posteriors are drawn from the same underlying distribution.  We conclude that there is strong evidence for a different mixture of primordial disks around GK stars than around M dwarfs.

\subsection{Origin of Planetesimal Disk Variation}
Direct observational constraints on the distribution of mass in planetesimal disks are lacking. Estimates of planetesimal disk structure can be made by comparison to the mass profile of dust and gas disks \citep[e.g.][]{Raymond14} or planetary systems \citep[e.g.][]{Weidenschilling77, Chiang13}. However, neither the dust and gas disks nor the final systems that form from them necessarily have the same surface density profile as their planetesimal disk. Planetesimal disks form within the gas/dust disk, but they are not coupled to its viscous evolution. Consequently, the evolution of their structure is distinct. Once the planetesimal disk forms, the final planetary system is built through the scattering and merger of planetesimals. These scattering events can lead to a significant redistribution of mass in the system thus erasing any direct connection between the surface density profiles of the initial and final states.

The structure of planetesimal disks can also be approached from a modeling standpoint. \citet{Moriarty15} simulated the growth of planetesimal disks from the accretion of inward drifting pebbles. They found that variations in the mass of the protoplanetary disk, disk metallicity or the duration of accretion lead to planetesimal disks with different masses and radial profiles. Thus the range of initial planetesimal disk structures required for this work may be a natural consequence of the expected variation among protoplanetary disks.

\subsection{Long-term evolution}
The growth of planetesimal disks into planetary systems is a relatively fast process in the inner disk. Most of the evolution of these systems occurs within the first million years with collision rates dropping off significantly after that. That is not to say that longer-term evolution of these systems is not important in determining the final state of planetary systems. Other works \citep[e.g.][]{Pu15, Volk15} have studied how late-term dynamical instabilities can potentially reduce the number of planets in a system. The applicability of these results to the systems formed in our simulations is unclear because the initial structure of our planetary systems (e.g. planet masses and spacings) differs from theirs. We extended a subset of 32 simulations out to about 5 billion years. Of the simulations that we extended, the observed planet multiplicity distribution did not change significantly between 10 million and 5 billion years. It is possible that a larger sample would result in more dynamical instabilities but unlikely at a high enough rate to account for the number of singly-transiting systems.

Tides are generally accepted to be an important factor in the orbital evolution of planets, particularly at small semi-major axes. Observations of exoplanets show an overall decrease in orbital eccentricity the closer the planet is to the star, which is attributed to tidal circularization. Tides would not have a large effect on our simulations because the timescale for tidal evolution is generally long compared to the duration of our simulations \citep{Jackson08b}. However, they would have a significant impact on the subsequent evolution of the orbits of the innermost planets in our simulations. In particular, the omission of tidal physics likely explains the increase in eccentricity with decreasing semi-major axis seen in some of our simulations, which is not observed. The tidal circularization of orbits is also accompanied by an inward migration of order 0.01-0.05 AU \citep{Jackson08b}. Both of these factors affect transit probability. Inward migration will increase the transit probability. However, orbital circularization will counteract this increase to some extent because eccentric orbits have a higher transit probability \citep{Burke08}.

\subsection{Implications and Predictions}

We note that the statistical comparisons made are not intended to ascertain the exact combination of initial conditions that must exist in planetesimal disks in order to produce the observed planetary systems. Rather, they are meant to emphasize the inability of models that assume all planetary systems have the same underlying architecture to reproduce the observations and, furthermore, to show that the range of system architectures that result from differing initial conditions in the planetesimal disk can reproduce important aspects of observations.

Nevertheless, we can draw some important conclusions from our analysis. To begin with, we, like other works \citep[e.g.][]{Hansen13}, find that a model for the final assembly of close-in planets that relies solely on the gravitational interactions of the growing bodies without the need for additional physics (e.g. planet disk interactions) can reproduce many of the observed characteristics of exoplanetary systems. Assuming that close-in planetary systems are predominantly gravitationally assembled in this way, we can say that some variation in the initial planetesimal disk structure is needed to form the variety of observed architectures (assuming there is no further relevant evolution of the systems after formation). In particular, some fraction of planetesimal disks must have initially steep radial surface density profiles while the rest have shallower surface density profiles.

The planetary systems that form from this range of initial conditions have significantly different final architectures. Because the systems that form from shallower surface density profiles typically only show at most one transiting planet and because these systems account for the majority of single tranet systems, we can make some general testable predictions about single tranet systems vs. multiple tranet systems:
\begin{itemize}
\item Single-tranet systems contain fewer, more widely separated (in terms of mutual Hill radii) planets than do multiple-tranet systems. Although not directly observable, this may be inferred by the frequency and strength of transit timing variations.
\item Single-tranet systems should show larger stellar obliquity. This phenomenon is observed by both \cite{Morton14} and \cite{Li15}.
\item Planet mass should increase with semi-major axis in single-tranet systems whereas in multiple tranet systems it should decrease with semi-major axis.
\item Single-tranet systems should have higher average eccentricity. This prediction is borne out in part by \cite{Limbach14}. They reported that average orbital eccentricity is a strong function of planetary multiplicity, with higher average eccentricity corresponding to fewer planets.
\end{itemize}

Additionally, because the diversity in architectures is determined at formation and is not the consequence of long-term evolution (e.g. dynamical instability), we predict that there should be no correlation in a system's multiplicity and its age.

\begin{figure*}
\begin{center}
\includegraphics[width=7in]{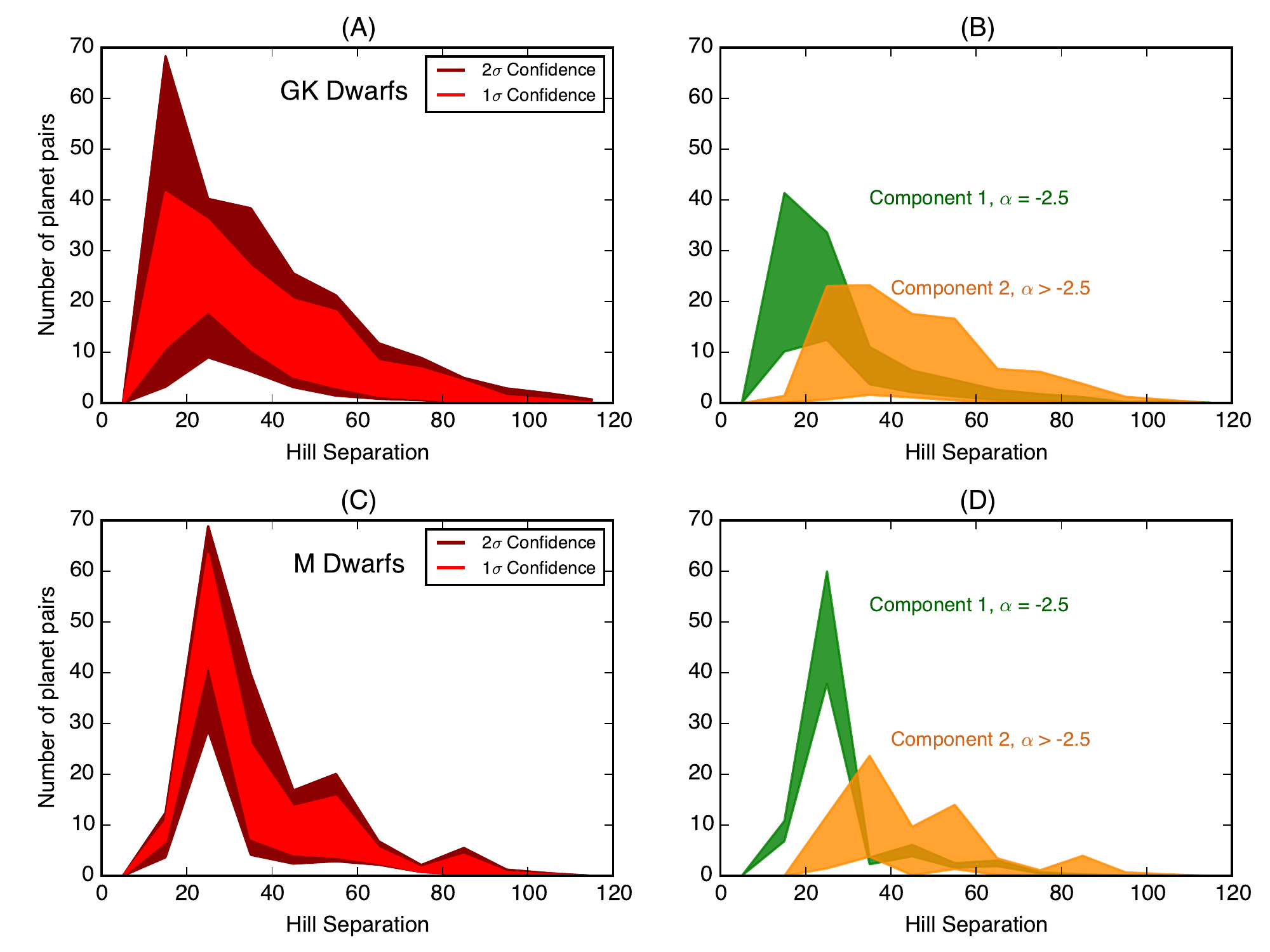}
\caption{Predictions for mutual Hill separation distributions, inferred from tranet fits for GK dwarfs (top row) and M dwarfs (bottom row). Red curves, at right, depict the model prediction at 1 and $2\sigma$ confidence in light and dark red, respectively. Contributions from each component of the mixture model at $1\sigma$ confidence indicated in orange (steeper surface density law) and green (shallower surface density power law).}
\end{center}
\label{fig:hill_separation}
\end{figure*}

In Figure \ref{fig:hill_separation}, we show our predicted distribution in the Hill separation of {\it detected} \kepler\ planets that we infer from the tranet comparison. The bulk of planet pairs have separations between $\sim$10 and 40 mutual hill radii. \citet{Pu15} find that the distribution of separations for high multiplicity systems (4+) tends to be condensed into the 10-25 range, slightly smaller values than our overall distribution. However, the component of our distribution corresponding to the majority of multi-tranet systems and all high multiplicity systems ($\alpha$=-2.5) tends to produce planet pairs with smaller separations more consistent with the \citet{Pu15} distribution (see Figure \ref{fig:hill_separation}b).

\section{Conclusions}
Recent analyses of exoplanet statistics have revealed that the configurations of planetary systems are not well described by a single-component model. Models that account for the frequency of multi-tranet systems significantly under-predict the number of single-tranet systems and models that can account for the number of single-tranet systems under-predict the number of multi-tranet systems. We have investigated the hypothesis that these differences in system architecture arise from variation in the primordial distribution of mass in planetesimal disks.

We performed a set of N-body simulations of late-stage planet formation, within which we varied initial planetesimal disk profiles. The resulting planetary systems are highly dependent on the initial conditions. Multiplicity, mutual inclinations, orbital separations, and eccentricities of planets vary significantly as a function of both total planetesimal disk mass and the radial distribution of this mass.

We compared our simulated systems to the \kepler\ yield of transiting planets. We apply the empirically-determined \kepler\ completeness to our synthetic populations of planets. In this way, we compare the outcome of our simulations to the observed multiplicities from \kepler. We conclude, similarly to previous works, that we cannot replicate the \kepler\  yield of exoplanets with a single model for planet occurrence. Rather, the data favor a mixture model, with at least two sets of distinct initial conditions required to replicate the observations from \kepler. We find that the same mixture model that correctly predicts the tranet distribution from \kepler\ also self-consistently reproduces the period ratio and duration ratio distributions.

We find that the best fit underlying population to our GK dwarf sample consists of one component accounting for approximately 20\% of systems that forms from a very steep initial planetesimal disk. This component accounts for most of the multi-tranet systems. The other component (accounting for approximately 80\% of systems) forms from planetesimal disks with shallower surface density profiles. Applying this analysis to M dwarfs, we see that the fraction of systems forming from steep planetesimal disk profiles increases to approximately 70\%. We conclude that the differing nature of the \kepler\ dichotomy between GK and M dwarfs indicates that the process of planet formation varies with stellar mass.

\acknowledgments
This material is, in part, based upon work supported by NASA under award No. NNX12AC01G and performed, in part, under contract with the California Institute of Technology (Caltech) funded by NASA through the Sagan Fellowship Program, and also by the MIT Torres Fellowship for Exoplanet Research. It is based on observations made with \kepler, which was competitively selected as the tenth Discovery mission. Funding for this mission is provided by NASA’s Science Mission Directorate. The authors would like to thank the many people who generously gave so much their time to make this Mission a success. This research used the NASA Exoplanet Archive, which is operated by the California Institute of Technology, under contract with the National Aeronautics and Space Administration under the Exoplanet Exploration Program. The simulations for this work were supported by the facilities and staff of the Yale University Faculty of Arts and Sciences High Performance Computing Center, and by the National Science Foundation under grant \#CNS 08-21132 that partially funded acquisition of the facilities. Lastly, we thank Eric Agol, Rory Barnes, Tom Quinn and many others for their valuable input and advice.

%\bibliography{mybib}

\end{document}